\documentclass[twocolumn,a4paper,showpacs,preprintnumbers,graphicx,reprint,amssymb, amsmath,aip,apl,floatfix]{revtex4-1}

\pdfoutput=1
\usepackage{docs}%
\usepackage{bm}%
\usepackage{graphicx,color}
\usepackage[colorlinks = true, citecolor = blue, linkcolor = red]{hyperref}

\begin{document}


\title{Mask aligner for ultrahigh vacuum with capacitive distance control} 

\author{Priyamvada Bhaskar}
\affiliation{II. Institute of Physics B and JARA-FIT, RWTH Aachen University, 52056 Aachen, Germany}
\email{bhaskar@physik.rwth-aachen.de}

\author{Simon Mathioudakis}
\affiliation{II. Institute of Physics B and JARA-FIT, RWTH Aachen University, 52056 Aachen, Germany}

\author{Tim Olschewski}
\affiliation{II. Institute of Physics B and JARA-FIT, RWTH Aachen University, 52056 Aachen, Germany}

 \author{Florian Muckel}
 \affiliation{II. Institute of Physics B and JARA-FIT, RWTH Aachen University, 52056 Aachen, Germany}

 \author{Jan Raphael Bindel}
 \affiliation{II. Institute of Physics B and JARA-FIT, RWTH Aachen University, 52056 Aachen, Germany}

\author{Marco Pratzer}
\affiliation{II. Institute of Physics B and JARA-FIT, RWTH Aachen University, 52056 Aachen, Germany}

 \author{Marcus Liebmann}
 \affiliation{II. Institute of Physics B and JARA-FIT, RWTH Aachen University, 52056 Aachen, Germany}

 \author{Markus Morgenstern}
 \affiliation{II. Institute of Physics B and JARA-FIT, RWTH Aachen University, 52056 Aachen, Germany}

\date{\today}

\begin{abstract}
We present a mask aligner driven by three piezo motors which guides and aligns a SiN shadow mask under capacitive control towards a sample surface. The three capacitors for read out are located at the backside of the thin mask such that the mask can be placed in $\mu$m distance from the sample surface, while keeping it parallel to the surface. Samples and masks can be exchanged in-situ and the mask can additionally be displaced parallel to the surface. We demonstrate an edge sharpness of the deposited structures below 100~nm, which is likely limited by the diffusion of the deposited Au on Si(111).
\end{abstract}


\maketitle 

Lateral nanostructuring is the base of a multifold of experimental research areas, such as mesoscopics, solid-state based quantum computing, nanoelectromechanical systems (NEMS), or photonics. Mostly, electron beam or optical lithography are employed, wherein the use of solvent-based resists inevitably leaves residues, which compromise the surface quality.
This is detrimental for surface science techniques in ultrahigh vacuum (UHV) such as scanning tunneling microscopy (STM).\\
An appealing alternative within UHV is the bottom-up assembly of individual atoms \cite{LYO1991,Eigler1990,Custance2009,Crommie1993}. It has been automated\cite{Celotta2014,Kalff2016} and has even been combined with mesoscopic device technology\cite{Vasko2011,Fuechsle2012}. The method is extremely precise at the atomic level \cite{Loth2012,Khajetoorians2013,Choi2017} and can be stable up to room temperature for certain types of manipulation schemes\cite{Shen1995}. However, it becomes increasingly tedious for more complex mesoscopic structures.\\
A faster alternative, which avoids chemical resists, is direct imprinting of structures by an atomic force microscope\cite{Garcia2014}. However, the corresponding lithography is restricted to selected materials and, typically, requires an environment which is not compatible with UHV. An example for the latter is local anodic oxidation\cite{Dagata1990}, a single-step nanolithography with sub-10\,nm resolution.\cite{Garcia2006} It has been applied, e.g., to fabricate mesoscopic structures in  GaAs \cite{Fuhrer2001} or graphene \cite{Puddy2013,Magda2014}.
However, since oxidation is the essential step, this technique lacks versatility in terms of materials.\\
Some UHV-based methods for nanostructuring are even more focused to a particular material as, e.g., the defect charging of BN below graphene \cite{Lee2016}, which recently enabled ultraclean graphene quantum dots of mesoscopic size to be probed by STM \cite{Ghahari2017}.\\
A  more versatile alternative for UHV nanostructuring is shadow mask evaporation, also called stencil lithography\cite{Du2017}. Silicon-based shadow masks have been produced with feature sizes less than 10\,nm\cite{Deshmukh1999} employing e-beam lithography\cite{Ono1996}, focussed ion beam lithography\cite{Matsui1996}, or, with larger throughput, ultraviolet optical lithography\cite{vandenBoogaart2004}. The central challenge beyond the mask production is to bring the mask close enough to the substrate in order to minimize the penumbra during evaporation.
The most simple solution is a rigid placement of the mask with respect to the substrate\cite{Gaertner2006}. However, here the mask-sample distance is either limited to tens of $\mu$m \cite{Stoeffler2015}, or the mask structure is in direct contact with the substrate\cite{Staley2007,Tien2016}. The latter has the drawbacks of possible additional contamination of the substrate \cite{Linklater2008} and difficulties to remove the mask in UHV. Another solution is based on the cantilever from a scanning force microscope (SFM). The SFM tip senses the surface while carrying the mask on top \cite{Luethi1999,Egger2005}. This led to structures, evaporated through the mask, with edge sharpness down to 10 nm \cite{Egger2005}, but comes with the instrumental overhead of a fully operational SFM. Finally, piezomotors have been utilized to move either the sample \cite{Zahl2005} or the mask \cite{Savu2008} in all three dimensions. However, so far, the distance calibration in the small distance range required an initial touching of the mask to the substrate, before the distance can be monitored, e.g., by a field emission current \cite{Steurer2014}. Again, an edge sharpness in the 10 nm range has been achieved, where diffusion of the deposited material is likely the limiting factor \cite{Tun2007,Linklater2008}.\\

Here, we present a capacitively controlled UHV mask aligner employing three piezo motors for approaching the mask to the substrate. Since the three capacitive sensors are placed directly on the backside of the 1\,$\mu$m thick SiN mask, the mask can be aligned with sub-$\mu$m precision relative to the substrate on a lateral scale of millimeters. This approach avoids an initial touching of the mask to the sample and requires less instrumental complexity than the SFM based technique. Lateral movement of the substrate relative to the mask (dynamical stencil lithography)\cite{Du2017}, is enabled by an additional, horizontal piezo motor. We show that sub-100 nm edge sharpness is possible with this UHV mask aligner, while edge sharpness partly goes down to 10 nm. Likely, both values are limited by the diffusion of Au on the oxidized Si(111) substrate.\cite{Linklater2008}



\begin{figure*}
	\includegraphics[width=16cm]{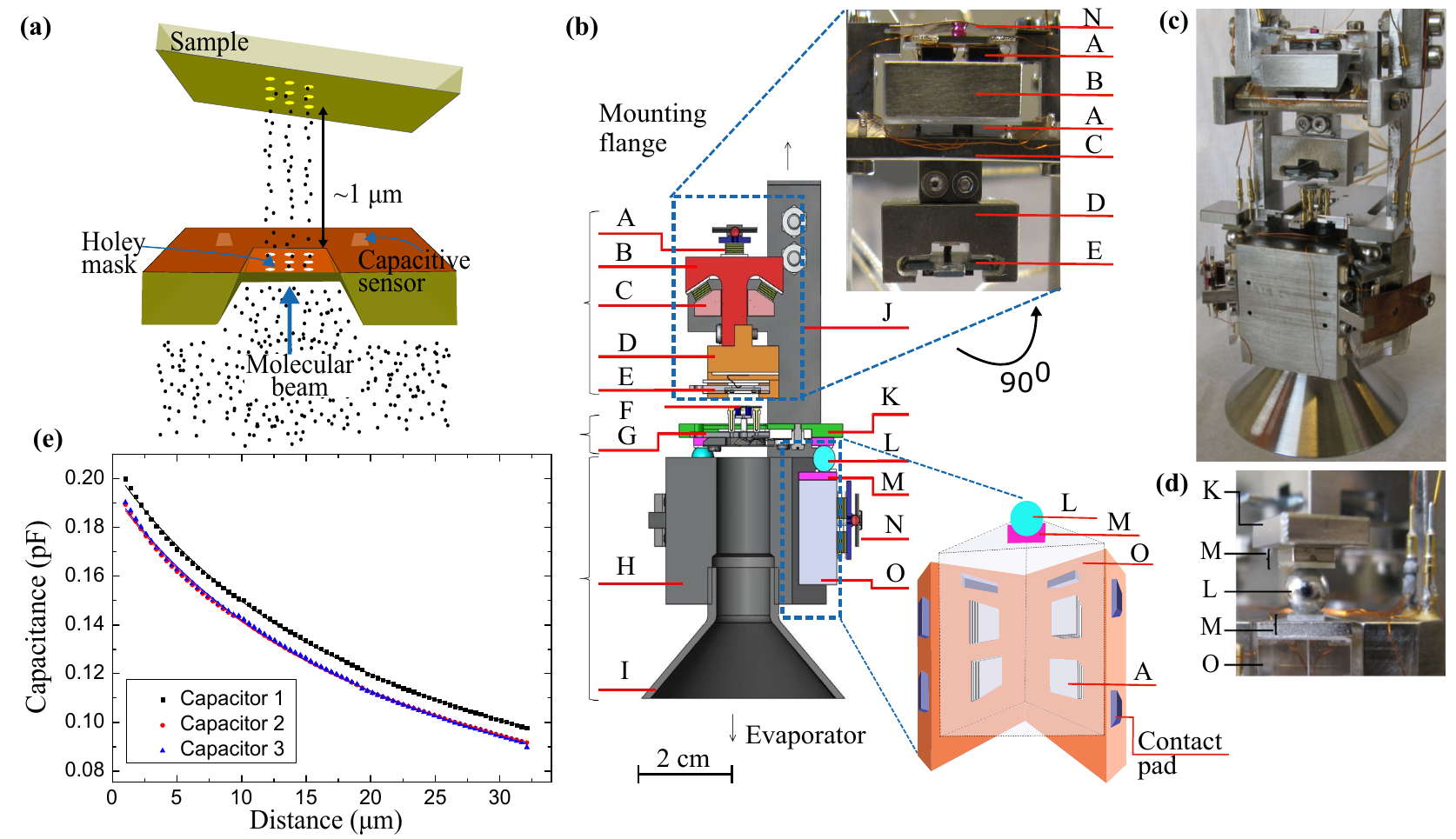}%
	\caption{\label{fig:1}(a) Sketch of mask and sample during molecular beam epitaxy (MBE) process (not to scale). Three capacitive sensors monitor the sample-mask distance, which enables parallel alignment of mask and sample. (b) Cross sectional drawing of the mask aligner (side view): (A) piezo-stack, (B) slider for horizontal sample movement, (C) sliding rail, (D) sample stage, (E) sample holder with sample, (F) shadow mask, (G) mask stage, (H) frame for piezomotors, (I) metal cone protecting the piezos from the molecular beam, (J) frame carrying the sample stage, (K) frame carrying the mask stage, (L) magnetic sphere, (M) two plates glued on top of sapphire prism (Al$_2$O$_3$, Ni), (N) pressure plate, (O) sapphire prism. Directions towards evaporator and mounting flange are marked by arrows. Top inset: optical image of the upper translation stage (front view) (same labels as in main image). Bottom inset: sketch of one of the three piezomotors (same labels as in main image). (c) Optical image of the assembled mask aligner (front view). (d) Optical image of the top part of a piezo motor portraying the flexible bearing of the mask frame via the magnetic sphere (L) (same labels as in (b)). (e) Capacitance curves of the three sensors during the mask approach to the substrate (points) with fit curves 
$C(d)=a+\frac{b}{c+d}$  (lines). }%
\end{figure*}
%

A geometrical estimate clarifies the required precision in distance control: For a distance $D$ between evaporation source and mask, distance $d$ between mask and substrate, and lateral extension of the evaporation source $W$, one gets a penumbra width of the evaporated structures:\cite{Linklater2008} $\Delta \approx\frac{d\cdot W}{D}$, assuming a ballistic path of the atoms from the evaporation source to the substrate. Hence, at typical $D=0.2$\,m  and $W=5\,$mm, as used in our experiment (see below), a penumbra width $\Delta = 10$\,nm requires $d=400$\,nm. A conservative estimate of the mean free path within the molecular beam $\lambda_{\rm MFP}$ is given by the value for an ideal gas\cite{reichl1980a,farrow1995molecular}
\begin{equation}
\lambda_{\rm MFP}=\frac{\sqrt{k_{\rm B}Tm_{\rm At}}}{2\pi^{3/2}R\rho\sigma^2} 
\end{equation}
at temperature $T$, deposition rate $R$, atomic mass $m_{\rm At}$, density of the deposited material $\rho$, and atomic scattering diameter $\sigma$.
Hence, $\lambda_{\rm MFP}$ is in the range of 100\,m at typical deposition rates of $R\approx 1$\,\AA/s, confirming the ballistic regime. Figure~\ref{fig:1}(a) shows a sketch of the mask and the sample. Three capacitive sensors are symmetrically placed around the mask with a mutual distance of 1\,mm. They enable high precision control of the distance to a conductive substrate, while adjusting the mask parallel to the sample surface.\\


\begin{figure*}
	\includegraphics[width=16cm]{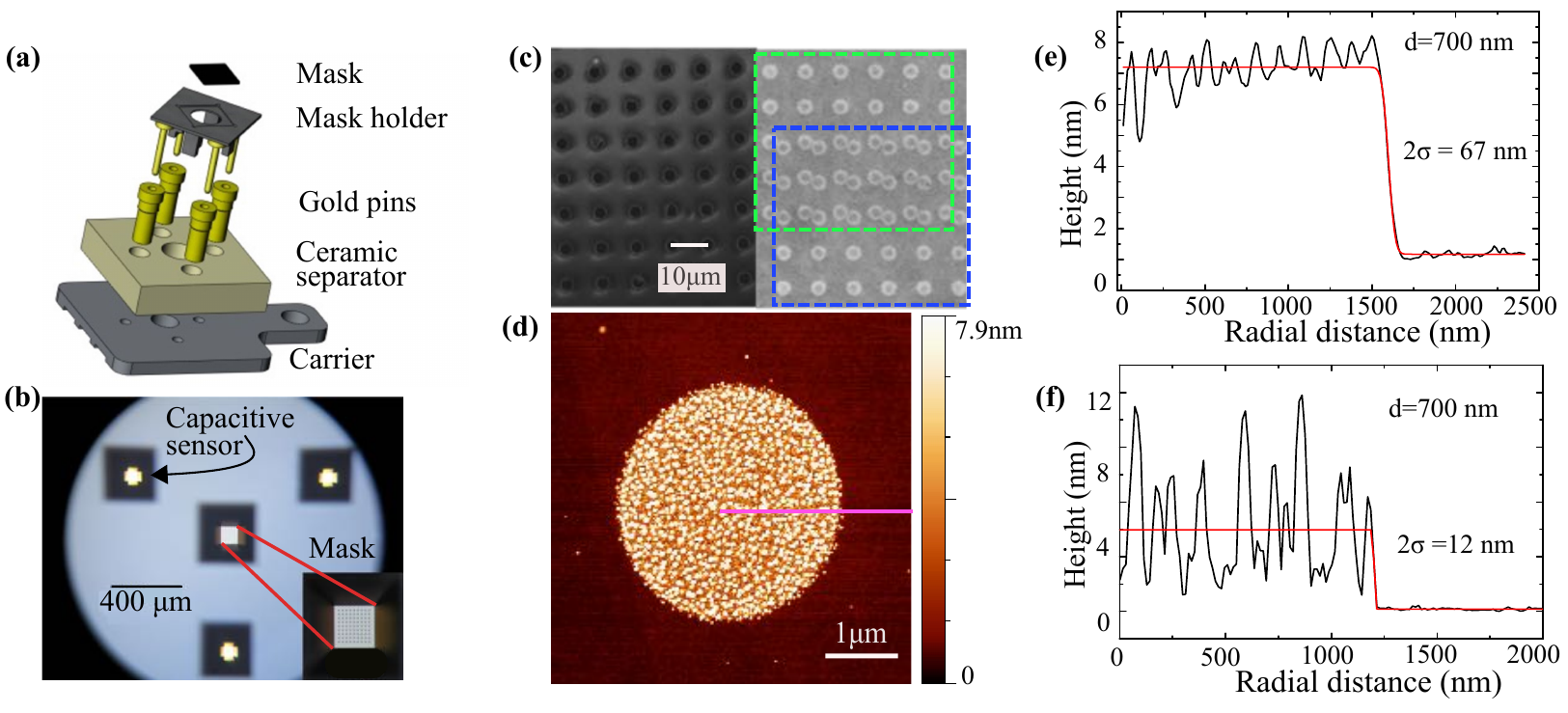}%
	\caption{\label{fig:2}(a) Exploded view of mask assembly including the shadow mask and the mask holder with four gold pins, which contact the capacitive sensors of the mask and ground. (b) Optical image of the shadow mask with the central mask region containing a holey pattern (inset) and three capacitive sensors. (c) SEM image of the holey Si$_3$N$_4$ mask (left) and of the Si(111) sample after evaporation of Au through the mask (right). Green and blue dashed rectangles indicate two subsequent evaporation steps with intermediate lateral displacement of the sample relative to the mask. (d) SFM image of one deposited Au circle at mask-sample distance $d=700$\,nm. The pink line marks the direction of a profile line. (e) Angularly averaged profile line at $d=700$\,nm. (f) Individual profile line at $d=700$\,nm. The red lines in (e) and (f) are fitted error functions to the experimental data (black lines) with indicated $2\sigma$ width.}%
\end{figure*}

The approach of the mask to the substrate is realized by three piezo motors. A cross section of the corresponding mask aligner is shown in Fig.~\ref{fig:1}(b). 
It consists of three modules made from stainless steel. The \textbf{lower motor module} (H, I, M, N, O) contains the three piezoelectric motors (lower inset) \cite{Pan1992,Panpatent1993} and enables mounting of the mask aligner to a CF flange. The piezomotors are assembled in three V-shaped recesses of the lower frame (H), being placed at a relative angle of $120^{\circ}$. Each motor consists of a sapphire prism (O) with an equilateral triangular base of 10\,mm edge length. The prism is clamped to the recess over four piezoelectric stacks (PI Ceramic GmbH) (A) covered each by a polished Al$_2$O$_3$ plate. Two further piezo-stacks are clamped to the front-end of the sapphire prism via a pressure plate (N). The pressure on this plate is adjusted via a screw, which strains the 0.2\,mm thick CuBe spring contacting the plate via a ruby ball (diameter: 2\,mm).\cite{Wiebe2004} Each prism, hence, can be moved via slip-stick motion employing a simultaneous saw-tooth voltage to all six piezo-stacks. Importantly, each of the three piezomotors can be moved independently. Step sizes during the movement mainly depend on the amplitude of the sawtooth voltage ($\sim 60$\,V) and the spring tension. They are typically in the range 50-200\,nm as measured by a ruler after $\sim 10^5$ consecutive steps. The maximum vertical displacement of each motor amounts to 9.5\,mm.\\
The \textbf{central mask module} (F, G, K) lies on top of the three prisms via three magnetic Nd spheres (L) (Fig.~\ref{fig:1}(d)). Each sphere is magnetically attached to the frame of the mask stage (K) and a prism (O) via Ni plates (M) glued to prism and frame. Small grooves in the Ni plates allow horizontal adjustment of the sphere. The sphere shape enables an inclination of the mask stage with respect to the individual piezomotors as required for the deliberate mask alignment. The frame (K) carries the mask stage, which allows exchange of
the mask via a standard sample holder of size $12\times 12$\,mm$^2$.
\\
Both frames (H, K) are pierced by a hole with a diameter of 12\,mm enabling the molecular beam to hit the mask.  A metal cone (I) prevents evaporation of material 
on the piezo-stacks and prisms.
\\
The \textbf{upper sample module} (A$-$E, J and upper inset) can move the sample parallel to the mask. It consists of a T-shaped stainless steel slider (B) holding the sample stage (D) and sliding via an additional piezomotor across a fixed sapphire rail (C). A horizontal range of 3.3\,mm is possible. Sample holder (E) and sample stage (D) are electrically isolated in order to reduce stray capacitances, since the sample surface acts as the counter electrode for the capacitive sensors on the mask (Fig.~\ref{fig:1}(a)). The sample can be exchanged in-situ via a standard sample holder of size $12\times 12$\,mm$^2$.\\
Typical capacitance curves $C(d)$ of the three sensors on the mask during an approach to a HOPG sample are shown in Fig.~\ref{fig:1}(e). They are measured employing an ac voltage while reading out the current response via a lock-in amplifier, leading to a sensitivity of $\sim 0.03$\,fF. The mask-sample distance $d$ is calibrated via the step size of the piezomotor (see above) and the contact point between mask and sample ($d=0$\,${\rm \mu}$m), which is taken as the point, where the $C(d)$ curve starts to saturate at small distances. The $C(d)$ curves of Fig.~\ref{fig:1}(e) can be nicely fitted by $C(d)=a+c/(b+d)$ with $a$, $b$, and $c$ being fit parameters. The deviation from the naively expected behavior $C(d)\approx \epsilon_{0}\epsilon_{\rm r}\frac{A}{d}$ ($A$: area of capacitor, $\epsilon_{0}$ ($\epsilon_{r}$): dielectric constant of vacuum (material)) is related to remaining stray capacitances, which also can exhibit a $d$ dependence. In order to minimize these stray capacitances, cables are shielded (also to avoid crosstalk) and contacts to the capacitive plates are provided from the backside. Since the capacitive sensors are ending at the backside of the 1\,$\mu$m thick Si$_3$N$_4$ mask, there is an additional vertical offset $d_0=1$\,$\mu$m in $C(d)$, which contributes to $b$.\\
Obviously, the capacitive control can start at distances $d > 30$\,$\mu$m, which can be easily determined optically within UHV using a long-distance microscope.\cite{Geringer2009} To compare the measured capacitance with the naively expected capacitance at $d=0$\,$\mu$m, we use the front end size of the capacitive sensor $A=50\times50$\,$\mu$m$^2$, the dielectric constant $\epsilon_{\rm r}=6-7$ of Si$_3$N$_4$, and the offset $d_0=1$\,$\mu$m, revealing $C(d=0)=140$\,fF. This is slightly smaller than the measured value of about $ 200$\,fF. Hence, stray capacitances still contribute. Indeed, we find variations in the $C(d=0)$ values for different masks by up to a factor of four. Nevertheless, distances corresponding to a single step of the piezomotor can always be detected reliably at $d<5$\,$\mu$m. Such a single step of one piezo motor corresponds to lifting of one of the capacitive sensors by $\sim 5$\,nm with respect to the others or to an angular misalignment of $\sim 0.0003^\circ$. This enables an extremely high precision of parallel alignment. To minimize the distance between mask and sample without touching, we approached ten different masks until contact showing that the gradient of the capacitance is a reliable measure, i.e., if ${\rm d}C/{\rm d}d \simeq 4$\,fF/$\mu$m, the mask does not touch and is closer than $d=3$\,$\mu$m to the
substrate. Note that the uncertainty of $3$\,$\mu$m in distance between the different capacitors still provides an angular precision of $\approx 0.2^\circ$.\\
%
%
%
Fig.~\ref{fig:2}(a) shows a sketch of the mask assembly enabling capacitive control and in-situ mask exchange. The mask carrier is a $12 \times 12 \times 1$\,mm$^3$ steel plate compatible with existing transfer mechanisms in UHV (Scienta Omicron). The mask structure ($3 \times 3 \times 0.2$\,mm$^3$) is glued to a mask holder carrying four gold pins at the underside, which are used to contact the capacitive sensor and ground. These gold pins are inserted into the four socket type gold pins mounted to the mask carrier, which  provide electrical contacts to springs in the mask stage when the carrier is inserted.\\
%
The top view of the shadow mask (Norcada Inc.) is shown in Fig.~\ref{fig:2}(b) and (c). It consists of a 1\,$\mu$m thick Si$_3$N$_4$ membrane of size (100\,$\mu$m)$^2$ carried by a highly doped Si wafer ($3-30$\,$\Omega$cm) of 0.2\,mm thickness. The membrane is produced by  depositing and structuring Si$_3$N$_4$ on top of the wafer and, subsequently, etching holes into the Si from the back. In the same etching step the three holes for the capacitors are produced which are covered by the same Si$_3$N$_4$ membrane, but without structure. Afterwards, 100\,nm of SiO$_2$ and 100\,nm of CrAu are deposited on the backside of these unstructured membranes. Each of the resulting thin capacitive CrAu plates is glued to a Au wire, which contacts to the gold pin of the mask holder. The inner membrane called mask in Fig.~\ref{fig:2}(b) is pierced by an array of $9\times 9$ holes with diameter 3\,$\mu$m (Fig.~\ref{fig:2}(c), left).\\ 

\begin{figure}
	\includegraphics[width=8.5cm]{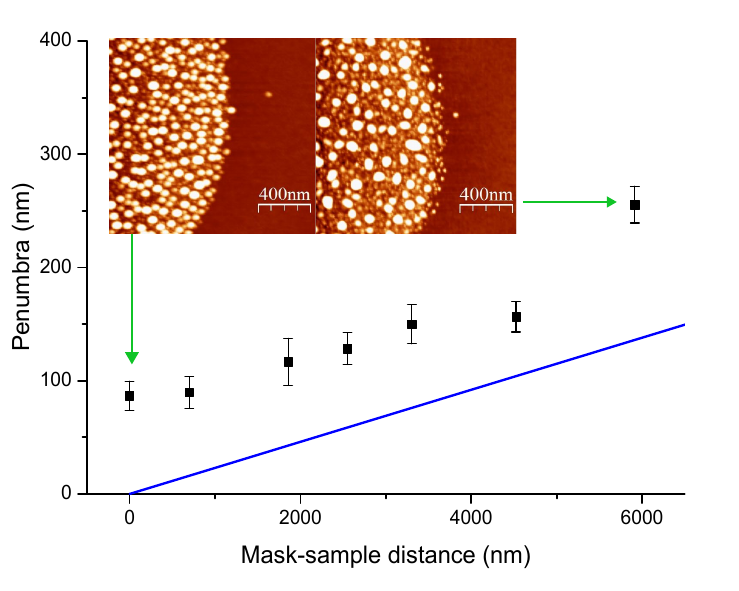}%
	\caption{\label{fig:3}2$\sigma$ widths of the error function fitted to angularly averaged profile lines (penumbra) of the deposited Au circles on Si(111) (black data points), $R=0.18$\,\r{A}/s, $T=300$\,K. The solid blue line gives the penumbra size expected geometrically for a ballistic path of the atoms during MBE growth. Insets: SFM images of the rims of two deposited circles corresponding to the data points at the green arrows.}
\end{figure}
Using this mask, we performed test evaporations of Au on a Si(111) sample at room temperature. The sample was prepared by ultrasonication in acetone and isopropyl alcohol, followed by oxygen plasma etching at 40\,W. The piezo motors are calibrated ex-situ revealing error bars in the step size of about 5 \%. Consecutive fields of $9\times 9$ circles are evaporated through the mask at a rate $R=0.18$\,\r{A}/s and at pressure $p=5\cdot 10^{-6}$\,Pa. Between the fields, the sample is moved closer to the mask and offset horizontally by $\sim 100$\,$\mu$m using the piezo motors. Figure~\ref{fig:2}c (right) displays a scanning electron microscopy (SEM) image of two overlapping fields, which have been used to crosscheck the steps of the horizontal piezo motor, revealing a discrepancy of only 4 \% with respect to the ex-situ calibration ($73\pm2$\,nm per step at amplitude 60\,V).\\
The evaporated circles are subsequently imaged by SFM. Figure~\ref{fig:2}(d) shows such a SFM image after evaporation at $d=700$\,nm. The circle with 7\,nm height consists of multiple clusters with diameters of $40-100$\,nm. It exhibits sharp edges, and a flat, clean surrounding, which reveals a successful transfer of the holey mask structure to the sample. The edge sharpness is basically determined by the cluster sizes (left inset in Fig.~\ref{fig:3}). Individual profile lines reveal an edge sharpness of $\sim 10$\,nm (Fig.~\ref{fig:2}(f)). To quantify the edge sharpness, we angularly averaged\cite{Horcas2007} the profile lines from the center of the disk towards its rim (pink line in Fig.~\ref{fig:2}(d)) and fitted an error function to them (Fig.~\ref{fig:2}(e)).
Note that although ellipticity is limited to a cluster size in our case, the angular average is very sensitive to possible lateral drifts of the mask relative to the sample, which would lead to an elliptical shape of the evaporated structures. \\ 
Figure \ref{fig:3} shows the 2$\sigma$ widths of the fitted error functions as a function of $d$. The error bar includes the variations between different circles, which are deposited simultaneously. The theoretical limit according to $\Delta =\frac{d\cdot W}{D}$ is provided for comparison (blue line). The offset of the data points with respect to the blue line is obviously close to the cluster size, implying that it is limited by diffusion. A diffusion induced broadening of $\sim 100$\,nm has recently also been found for Au evaporation on Si(001) at room temperature , after pressing a shadow mask directly onto the sample \cite{Linklater2008}. We note that even at the safe approach distance of $d\approx 3$\,$\mu$m, the penumbra size is $\Delta \sim 100$\,nm.\\
In conclusion, we described a novel type of UHV mask aligner employing piezo motors and a capacitive control for mask-sample alignment. We demonstrate edge sharpness down to 10\,nm and edge precision below 100\,nm, possibly limited by diffusion of Au on Si(111) at room temperature. The versatility of the mask aligner is given by the in-situ mask and sample exchange and the lateral movement of the sample relative to the mask.


\bibliographystyle{aipnum4-1}
\providecommand{\noopsort}[1]{}\providecommand{\singleletter}[1]{#1}%

\end{document}